\documentclass[conf]{new-aiaa}

\usepackage[utf8]{inputenc}
\usepackage{upgreek}

\usepackage{mathtools}
  \mathtoolsset{showonlyrefs, centercolon, mathic}
\usepackage{amssymb}
\usepackage{dsfont}
\usepackage{nicefrac}
\usepackage{siunitx}

\usepackage{siunitx}
\usepackage{longtable,tabularx}
\usepackage{booktabs,colortbl}
\usepackage{pgfplotstable}
  \pgfplotstableset{
    every even row/.style={before row={\rowcolor[gray]{0.9}}},
    every head row/.style={before row=\toprule,after row=\midrule},
    every last row/.style={after row=\bottomrule},
  }

\newcommand{\inputtikzpicture}[1]{%
  \includegraphics{{#1.tikz}.pdf}%
}

\newcommand{\trans}{^{\mathsf{T}}} 
\newcommand{\inv}{^{-1}} 
\newcommand{\ml}{_{\ensuremath{\mathrm{ML}}}} 

\DeclareMathOperator*{\argmax}{arg\,max}

\DeclareMathOperator*{\maximize}{maximize}
\DeclareMathOperator*{\minimize}{minimize}
\DeclareMathOperator*{\subjectto}{subject\ to}

\newcommand{\dd}{\ensuremath{\mathrm{d}}} 
\newcommand{\reals}{\ensuremath{\mathds{R}}} 

\newcommand{\nx}{{\ensuremath{n_{\mathrm{x}}}}} 
\newcommand{\ny}{{\ensuremath{n_{\mathrm{y}}}}} 
\newcommand{\ninp}{{\ensuremath{n_{\mathrm{u}}}}} 
\newcommand{\npar}{{\ensuremath{n_{\theta}}}} 
\newcommand{\ns}{{\ensuremath{n_{\mathrm{s}}}}} 

\newcommand{\pe}{\ensuremath{p_{\mathrm e}}} 
\newcommand{\pz}{\ensuremath{p_{\mathrm z_{1:N}}}} 
\newcommand{\elle}{\ensuremath{\ell_{\mathrm e}}} 
\newcommand{\ellz}{\ensuremath{\ell}} 

\newcommand{\tm}{{\ensuremath{\tilde t}}} 

\newcommand{\yp}{\ensuremath{\hat y}} 
\newcommand{\xp}{\ensuremath{\hat x}} 

\newcommand{\im}{\ensuremath{\mathcal I}} 

\newcommand{\Vref}{\ensuremath{V_{\operatorname{ref}}}}

\hypersetup{colorlinks, linkcolor=blue, citecolor=blue, urlcolor=blue,}

\title{Collocation-Based Output-Error Method for Aircraft System Identification}

\author{Dimas Abreu Archanjo Dutra%
  \footnote{%
    Adjunct Professor, Departamento de Engenharia Mecânica,
    Av. Pres. Antônio Carlos, 6627, Member AIAA.
  }
}
\affil{%
  Universidade Federal de Minas Gerais, Belo Horizonte, Minas Gerais, Brazil,
  31.270-901
}

\begin{document}

\maketitle

\begin{abstract}
  The output-error method is a mainstay of aircraft system identification
  from flight-test data. It is the method of choice for a wide range of 
  applications, from the estimation of stability and control derivatives for 
  aerodynamic database generation to sensor bias estimation in flight-path 
  reconstruction.
  However, notable limitations of the output-error method are that it requires
  ad hoc modifications for applications to unstable systems and it is an
  iterative method which is particularly sensitive to the initial guess.
  In this paper, we show how to reformulate the estimation as a collocation
  problem, an approach common in other disciplines but seldomly used in flight
  vehicle system identification.
  Both formulations are equivalent in terms of having the same solution, but
  the collocation-based can be applied without modifications or loss of 
  efficiency to unstable systems.
  Examples with simulated and real-world flight-test data also show that
  convergence to the optimum is obtained even with poor initial guesses.
\end{abstract}

\section*{Nomenclature}

\subsection*{Mathematical symbols}
{\renewcommand\arraystretch{1.0}
\noindent\begin{longtable*}{@{}l @{\quad} l@{}}
$\reals$ & the set of real numbers \\
$t$ & time \\
$A\inv$ & matrix inverse of $A$ \\
$A\trans$ & matrix (or vector) transpose of $A$ \\
$\dot x$ & time derivative of $x$, i.e., $\nicefrac{\dd x}{\dd t}$ \\
$C_D$, $C_C$, $C_L$  & wind-axis force coefficients \\
$C_\ell$, $C_m$, $C_n$  & body-axis moment coefficients \\
$\nx$, $\ny$ & number of system states and outputs \\
$\ninp$, $\npar$ & number of system inputs and unknown parameters \\
$N$ & number of measurement instants in the experiment \\
$:=$ & equal by definition, i.e., defined as \\
$\ln$ & natural logarithm (base $e$) \\
$\det$ & matrix determinant \\
\end{longtable*}}

\subsection*{Acronyms}
{\renewcommand\arraystretch{1.0}
\noindent\begin{longtable*}{@{}l @{\quad} l@{}}
CEA & Centro de Estudos Aeronáuticos \\
CEACoEst & CEA Control and Estimation Library \\
CPU & Central Processing Unit. \\
COIN-OR & Computational Infrastructure for Operations Research. \\
DLR & German Aerospace Center. \\
FAA & Federal Aviation Administration.\\
GPU & Graphics Processing Unit. \\
HARV & High Angle of Attack Research Vechicle. \\
HATP & High Angle of Attack Technology Program. \\
i.i.d. & Independent and Identically Distributed. \\
IPOPT & Interior Point OPTimizer. \\
IVP & Initial-Value Problem. \\
ML & Maximum Likelihood. \\
ODE & Ordinary Differential Equation. \\
OEM & Output-Error Method. \\
SIMD & Single Instruction, Multiple Data. \\
\end{longtable*}}

\section{Introduction}
\lettrine{A}{s} the need for building mathematical models of dynamical systems
from data, known as \emph{system identification}, arises in several application
areas, several scientific communities are active in the development of its
theory and algorithms, as noted by Ljung \cite[Sec.~3]{ljung2010psi} in a 
recent survey.
The different uses of the identified models and the context of the systems
and tests shapes the identification methods.
In the systems and automatic control community, for example, a major need
for system identification was the obtention of models for control
\cite[Sec.~3.9]{ljung2010psi}.
As in their main application only the input--output relationship matters,
a focus can be seen on black-box, discrete-time, transfer function models
and methods 
\cite{ljung2010psi, astrom1965nil, astrom1971sis, ljung1999si}.

For aircraft system identification, however, a major driver were  research
aircraft with novel configurations or operating at previously unexplored
flight regimes.
Notable examples of novel configurations are the lifting-body M2 prototypes
and HL-10, tilt-wing XC-142A, thrust-vectored X-31, human-powered Gossamer
Albatross, and the active aeroelastic wing program.
Among previously unexplored flight regimes we have hypersonic speeds as in the
X-15 and the space shuttle; and high angles of attack which were studied in
the High Angle of Attack Technology Program (HATP), the F-18 High Angle of
Attack Research Vechicle (HARV), and in the X-29.
For more details on these aircraft and the role of system identification in 
their programs, consult the reviews and retrospectives of the area in 
Refs.~\citenum{klein1989eaa, hamel1996efv, wang2004rre, morelli2005asi,
jategaonkar2004ams} and the ample references therein.
The investigative nature of these programs favored parameter estimation 
techniques that allowed for a deeper understanding of the underlying 
aerodynamic phenomena, their relationship to the aircraft stability and 
control, and the validity of the theoretical and wind-tunnel results.
System identification was also a supporting technique in the safe expasion of 
the operational flight envelope.

To some extent, these needs are also present in the flight-testing of production
aircraft prototypes, for which system identification is also routinely used 
\cite{klein2006asi, jategaonkar2015fvs}.
For these, the estimated stability and control derivatives are also used to
build high-fidelity aerodynamic databases for training, engineering and
in-flight simulators meeting the FAA requirements.
Finally, the identified models are also used for control law synthesis and
controller tuning.
Consequently, for flight vehicle system identification the focus has been on
white-box or light-gray-box phenomenological models, whose parameters have 
physical meaning and can be related across different reference flight 
conditions.
Most models are multiple-input multiple-output and are represented in
the state-space form.

With respect to the estimation algorithms, for aircraft the maximum likelihood
(ML) estimators have dominated.
Of these, the output-error method (OEM), which amounts to a nonlinear weighted
least-squares problem when the noise covariance matrix is assumed known is
the most widely used \cite[p.~681]{jategaonkar2004ams}, even for applications
such as flight-path reconstruction \cite{mulder1999nla} and aeroelastic 
modeling \cite{silva2011dgp, silva2012sif}.
Notable limitations of the classical OEM, however, are its inapplicability to
unstable systems [\citenum{jategaonkar1994epe}; \citenum{jategaonkar2015fvs},
Chap.~9] and its convergence to local minima when poor initial estimates 
are used.

A field with similar requirements for system identification as for flight 
vehicles is chemical engineering, in which the output error method was also 
developed independently%
\footnote{%
  The author's search did not yield any references to the aircraft literature 
  when the OEM was developed in the chemical engineering literature.
}
\cite{swartz1975dpe, bock1981nti, bock1983rap, schloder1983irc, lohmann1992nmp}.
Their applications also required estimation of parameters of white-box 
continuous-time models in state-space, usually reaction rates in chemical
equations.
The application of the maximum likelihood principle to this class of problems
leads, naturally, to the output-error method.
Differences in their application context, however, resulted in alternative
implementations routes.
Some estimation problems in chemical engineering involve 
ill-conditioned differential equations which may not have a valid solution
for some parameter values.
Additionally, not much a-priori information is available on the parameters for
the initial guess of iterative methods.
This should be contrasted with aircraft parameter estimation, for which
reasonably good guesses for the stability and control derivatives can be 
obtained from theoretical analyses, wind-tunnel experiments and 
computational-fluid-dynamics calculations.

As a result, the chemical reaction system identification community developed
implementations of the OEM capable of coping with these issues: multiple
shooting and collocation \cite{bock1981nti, bock1983rap}.
In these approaches to implement the output-error method, the decision
variables of the optimization problem are augmented with the states at several
mesh points across the experiment time interval.
Equality constraints are then added to encode the system dynamics and enforce
continuity of the solution.
These methods are also very common in optimal control 
\cite{betts2010pmo}, as there is a certain duality between estimation
and control, and were developed by researchers active in both fields
like H.~G.~Bock \cite{bock1984msa}.
The author has also used collocation methods in state and parameter estimation
problems \cite{dutra2014map, dutra2017jmp}.

Traditionally, in the aeronautical literature, the output-error method is 
implemented using the single shooting or initial-value problem (IVP) approach,
in which only the
unknown parameters and initial values of the states are used as decision
variables of the nonlinear optimization problem.
Good explanations of the problems that arise with this formulation were given
by Bock [\citenum{bock1981nti}, Sec.~8.2; \citenum{bock1983rap}, p.~97]:
neglecting the information of the states by rewriting the problem only in terms
of the parameters results in a ``reinversion of the inverse problem'' which
spoils ``the very nature of the inverse problem'', leading to a ``deterioration
of efficiency'' and a ``substantial loss of stability''.
A defining property of the system identification problems is that there is
plentiful information on the states but comparatively little on the parameters.
This information is efficiently exploited by estimation methods
implemented with the collocation and multiple-shooting approaches.

A notable improvement allowed by collocation and multiple shooting to aircraft
system identification with the output-error method is applicability to unstable
systems without ad hoc modifications.
The multiple shooting formulation has already been applied to unstable flight
vehicles for that end [\citenum{jategaonkar1994epe}; 
\citenum{jategaonkar2015fvs}, Sec.~9.12], but is generally shunned in favor
of the stabilized OEM due to programming complexity and reuse of existing
code \cite[Sec.~9.15]{jategaonkar2015fvs}.
Another disadvantage of the single shooting problem formulation with respect to
the state-augmented ones is divergence of the optimization and convergence to
suboptimal local minima.
While these issues may not be a problem for estimation of the rigid-body
dynamics, they are an issue for
aeroelastic or aeroservoelastic system identification which feature larger
and more complex models with parameters that are more difficult to predict
accurately \cite[Sec.~I]{grauer2018rtp}.

In this paper, we present the collocation formulation of the output-error 
method, which is seldomly used for system identification of aircraft from
flight-test data---one of the few examples available was presented by
\citet[Sec.~7.4]{betts2010pmo}.
The method is applied to experimental and simulated flight data bundled with the
supplemental material of \citet{jategaonkar2015fvs} and ground vibration test
data from the experiment described by \citet{gupta2016gvt}.
The convergence basin of the collocation implementation is explored to 
investigate its robustness against poor starting values for the parameters.

This paper is organized as follows.
In Sec.~\ref{sec:oem}, the output-error method is presented, together with the
three main approaches to implement it: single shooting, multiple shooting
and collocation.
In Sec.~\ref{sec:examples} the properties of the collocation implementation
are illustrated with three examples.
Finally, in Sec.~\ref{sec:conclusion} the conclusions are drawn out and 
directions for future work are outlined.

\section{The output-error method}
\label{sec:oem}
In this section we formulate the output-error method for parameter estimation
and describe the three main implementations used to write it as a 
nonlinear optimization problem in a form suitable for computational solution.

\subsection{Problem formulation}
\subsubsection{General case}
The general problem consists of estimating the vector of unknown parameters
$\theta\in\reals^\npar$ of a system 
described by ordinary differential equations (ODEs) of the form
\begin{align}
  \label{eq:state_dynamics}
  \dot x(t) &= f\big(x(t), u(t), \theta\big), &
  x(0) &= x_0(\theta),\\
\end{align}
where $t\in\reals$ is the time, $x\colon\reals\to\reals^\nx$ is the 
state-path of the system, $u\colon\reals\to\reals^\ninp$ are the external
inputs,
$f\colon\reals^\nx\times\reals^\ninp\times\reals^\npar\to\reals^\nx$ is
a possibly nonlinear function encoding the system dynamics, and 
$x_0\colon\reals^\npar\to\reals^\nx$ is the initial state of the experiment
which is, without loss of generality, assumed to start at $t=0$.
The parameter vector $\theta$ holds all unknown system parameters, which may
include aerodynamic stability and control derivatives, initial states, sensor
bias, and even parameters of the stochastic noise model such as variance.
The system outputs $y\colon\reals\to\reals^\ny$ are described by a static
function $g\colon\reals^\nx\times\reals^\ninp\times\reals^\npar\to\reals^\ny$
which can also depend on the unknown parameters,
\begin{align}
  \label{eq:output_equations}
  y(t) &= g\big(x(t), u(t), \theta\big).
\end{align}

While the outputs $y$ represent the ideal measurables, we only have access
to the measurements $z_k$, sampled at a finite number $N$ of time instants 
$\tm_k$.
Under the statistical framework of maximum likelihood estimation, the $z_k$
are interpreted as noise-corrupted measurements of the outputs, 
\begin{align}
  \label{eq:z_noise_interpret}
  z_k &= y(\tm_k) + e_k,
\end{align}
where the measurement noise $e_k$ is described by some suitable probability
distribution.
The maximum likelihood estimate $\hat\theta\ml$ is then the value which
maximizes the likelihood function: the joint conditional probability density
$\pz$ of all measurements, given the parameters
\begin{align}
  \label{eq:maximum_likelihood}
  \hat\theta\ml &
  := \argmax_\theta \pz\big(z_1,\dotsc,z_N \big|\theta\big).
\end{align}
Under the assumption that the noise is idenpendent and identically distributed
(i.i.d.) across measurement instants, Eq.~\eqref{eq:maximum_likelihood} can be
simplified by noting that
\begin{align}
  \pz\big(z_1,\dotsc,z_N \big|\theta\big) = 
  \prod_{k=1}^N \pe\big(z_k - \yp(\tm_k, \theta) \big| \theta \big),
\end{align}
where $\pe(\cdot|\theta)$ is the conditional probability density of the
noise vectors $e_k$ and $\yp(t,\theta)$ is the predicted system output
for a given value of  $\theta$, i.e., the solution of
Eqs.~\eqref{eq:state_dynamics} and \eqref{eq:output_equations},
\begin{subequations}
  \begin{align}
    \label{eq:pred_x}
    \dot\xp(t,\theta) &= f\big(\xp(t,\theta), u(t), \theta\big), \qquad
    \xp(0,\theta) = x_0(\theta),\\
    \label{eq:pred_y}
    \yp(t,\theta) &:= g\big(\xp(t,\theta), u(t), \theta\big).
  \end{align}
\end{subequations}

As the logarithm is a monotonic function which does not change the location
of maxima, the logarithm of the likelihood function, known as the 
log-likelihood, is usually maximized in 
\eqref{eq:maximum_likelihood} as it leads to better-conditioned problems:
\begin{subequations}
  \begin{align}
    \label{eq:ellz_defn}
    \ellz(\theta) &:= \ln \pz(z_1,\dotsc,z_N |\theta) 
    =\sum_{k=1}^N\ln \pe\big(z_k - \yp(\tm_k, \theta) \big| \theta\big) 
    = \sum_{k=1}^N \elle(z_k - \yp(\tm_k, \theta), \theta),\\
    \label{eq:elle_defn}
    \elle(\epsilon, \theta) &:= \ln \pe(\epsilon | \theta).\\
  \end{align}
\end{subequations}
The most general form of the output-error method estimation problem is then
\begin{subequations}
  \label{eq:ml_loglike}
  \begin{alignat}{3}
    \label{eq:ml_loglike_obj}
    \maximize_{\theta\in\reals^\npar} & \quad&&
    \sum_{k=1}^N \elle(z_k - \yp(\tm_k, \theta), \theta)\\
    \label{eq:ml_loglike_cinit}
    \subjectto &&&
    \xp(0,\theta) = x_0(\theta) \\
    \label{eq:ml_loglike_cxdot}
    &&&
    \dot\xp(t,\theta) = f\big(\xp(t,\theta), u(t), \theta\big) \\
    \label{eq:ml_loglike_cy}
    &&&\yp(t,\theta) = g\big(\xp(t,\theta), u(t), \theta\big).
  \end{alignat}
\end{subequations}%
\noeqref{eq:ml_loglike_cxdot,eq:ml_loglike_cy, 
         eq:ml_loglike_cinit,eq:ml_loglike_obj}

\subsubsection{Special case: Gaussian measurement noise}
Due to the tractability of the maximization problems it produces and
its ubiquity in statistical models, the noise is often assumed to be zero-mean
and normally distributed.
In that case, denoting by $R\colon\reals^\npar\to\reals^{\ny\times\ny}$
its covariance matrix,
\begin{align}
  \label{eq:gaussian_noise}
  e_k \sim \mathcal{N}\big(0, R(\theta)\big)\qquad k=1,\dotsc, N,
\end{align}
the noise log-density is
\begin{align}
  \label{eq:pe_gaussian}
  \elle(\epsilon,\theta) &
  = -\tfrac12 \epsilon\trans R(\theta)\inv \epsilon 
  -\tfrac12\ln\det\big(R(\theta)\big)
  -\tfrac{\ny}{2}\ln(2\pi).
\end{align}
When the noise covariance is known or fixed ($R$ does not depend on $\theta$),
the last two terms in the right-hand side of \eqref{eq:pe_gaussian}
can be dropped as they are constants which do not influence the location of
maxima.
The maximum likelihood estimator reduces to a nonlinear weighted least-squares
problem with $R\inv$ as the weighting matrix,
\begin{subequations}
  \label{eq:ml_gauss_known_cov}
  \begin{alignat}{3}
    \label{eq:ml_gauss_known_cov_obj}
    \minimize_{\theta\in\reals^\npar} & \quad&&
    \sum_{k=1}^N 
    \big[ z_k - \yp(\tm_k, \theta)\big]\trans
    R\inv \big[ z_k - \yp(\tm_k, \theta)\big] \\
    \label{eq:ml_gauss_known_cov_cinit}
    \subjectto &&&
    \xp(0,\theta) = x_0(\theta) \\
    \label{eq:ml_gauss_known_cov_cxdot}
    &&&
    \dot\xp(t,\theta) = f\big(\xp(t,\theta), u(t), \theta\big) \\
    \label{eq:ml_gauss_known_cov_cy}
    &&&\yp(t,\theta) = g\big(\xp(t,\theta), u(t), \theta\big),
  \end{alignat}
\end{subequations}%
\noeqref{eq:ml_gauss_known_cov_cxdot,eq:ml_gauss_known_cov_cy, 
         eq:ml_gauss_known_cov_cinit,eq:ml_gauss_known_cov_obj}%
in which the sign of $\ell$ was inverted and the maximization
converted to a minimization. 
If, furthermore, the elements of the noise vector $e_k$ are assumed 
to be independent of each other, then $R$ is diagonal,
\begin{equation}
  \label{eq:r_diag}
  R =
  \begin{bmatrix}
    r_1 & 0 & \cdots & 0 \\
    0 & r_2 & \cdots & 0\\
    \vdots & \vdots & \ddots & \vdots\\
    0 & 0 & \cdots & r_\ny
  \end{bmatrix}
\end{equation}
and the maximum likelihood estimation problem is 
\begin{subequations}
  \label{eq:ml_gauss_known_cov_diag}
  \begin{alignat}{3}
    \label{eq:ml_gauss_known_cov_diag_obj}
    \minimize_{\theta\in\reals^\npar} & \quad&&
    \sum_{k=1}^N 
    \sum_{i=1}^\ny 
    \frac{\big[z_k - \yp(\tm_k, \theta)\big]^2_{i}}{r_i} \\
    \label{eq:ml_gauss_known_cov_diag_cinit}
    \subjectto &&&
    \xp(0,\theta) = x_0(\theta) \\
    \label{eq:ml_gauss_known_cov_diag_cxdot}
    &&&
    \dot\xp(t,\theta) = f\big(\xp(t,\theta), u(t), \theta\big) \\
    \label{eq:ml_gauss_known_cov_diag_cy}
    &&&\yp(t,\theta) = g\big(\xp(t,\theta), u(t), \theta\big),
  \end{alignat}
\end{subequations}%
\noeqref{eq:ml_gauss_known_cov_diag_cxdot,eq:ml_gauss_known_cov_diag_cy, 
         eq:ml_gauss_known_cov_diag_cinit,eq:ml_gauss_known_cov_diag_obj}%
in which $[\epsilon]_i$ denotes the $i$-th element of the $\epsilon$
vector.

\subsubsection{A pragmatic view on the statistical model}
We note that although the maximum likelihood estimator is derived from a
statiscal framework for which it is an optimal solution, it makes sense
regardless of the probabilistic iterpretation
\cite[see][pp.~16,~222]{ljung1999si}.
From a pragmatic viewpoint, we could regard the noise log-density $\elle$
in \eqref{eq:ml_loglike_obj} as a metric or norm on the error size.
In \eqref{eq:ml_gauss_known_cov_obj} and \eqref{eq:ml_gauss_known_cov_diag_obj},
for example, the estimate minimizes the
mean squared Mahalanobis distance between the measurements and the
model outputs.
When the log-density $\elle$ depends on the parameter $\theta$ as in
\eqref{eq:pe_gaussian}, the ``best'' metric is not known beforehand 
\cite[p.~200]{ljung1999si} so it is chosen by the help of a criterion which
penalizes ``small'' metrics, e.g., the $\ln\det\big(R(\theta)\big)$ 
term of \eqref{eq:pe_gaussian}.

The choice of the output noise distribution $\pe$ can then be transformed to
the choice of a metric $L\colon\reals^\ny\times\reals^\npar\to\reals$ for
the error size in the optimization problem 
\begin{subequations}
  \label{eq:metric_est}
  \begin{alignat}{3}
    \label{eq:metric_est_obj}
    \minimize_{\theta\in\reals^\npar} & \quad&&
    \sum_{k=1}^N 
    L( z_k - \yp(\tm_k, \theta), \theta) \\
    \label{eq:metric_est_cinit}
    \subjectto &&&
    \xp(0,\theta) = x_0(\theta) \\
    \label{eq:metric_est_cxdot}
    &&&
    \dot\xp(t,\theta) = f\big(\xp(t,\theta), u(t), \theta\big) \\
    \label{eq:metric_est_cy}
    &&&\yp(t,\theta) = g\big(\xp(t,\theta), u(t), \theta\big).
  \end{alignat}
\end{subequations}%
\noeqref{eq:metric_est_cxdot,eq:metric_est_cy, 
         eq:metric_est_cinit,eq:metric_est_obj}%
The estimation yields the ``smallest'' mean error, acording to $L$.
Several metrics have emerged as robust alternatives to the quadratic form
associated with the normal distribution
\cite{%
  aravkin2012nrs, aravkin2012rtf, aravkin2012sct, 
  aravkin2013spe,aravkin2017gks%
}.

\subsection{Single shooting implementation}

The optimization problems in the preceding section, represented in 
\eqref{eq:ml_loglike}, \eqref{eq:ml_gauss_known_cov}, 
\eqref{eq:ml_gauss_known_cov_diag}  and \eqref{eq:metric_est}, are
still in an abstract form, as the true solution of the ODE
\eqref{eq:state_dynamics} is referenced.
Except for the case of linear or affine systems, the predicted state-path does
not admit closed-form solutions.
Hence, the most obvious way to implement the output-error method is to compute
an approximate solution of the system state-path by using a numerical
integration scheme like a Runge--Kutta method.
This is known as the
single shooting or initial-value problem (IVP) implementation.
For a survey of numerical integration methods as it pertains to aircraft
system identification using single shooting, see 
\citet[Sec.~3.8.1]{jategaonkar2015fvs}.

A single time mesh of $m$ distinct instants $\tau_k$ of the experiment
interval $[0,T]$, including the endpoints, is selected to perform the 
integration:
\begin{equation}
  \label{eq:single-shooting_mesh}
  0 = \tau_1 < \tau_2 < \cdots < \tau_m = T.
\end{equation}
We will assume here that the mesh includes all measurement times $\tm_k$, 
as this simplifies the implementation and notation.
However, this is not needed if an associated interpolation scheme (dense output)
is used with the integration method \citep[e.g.,][]{shampine1985irk}.
In any case, 
it is important that the mesh be kept fixed during the optimization as its 
variation for different $\theta$, which can occur when using variable-step
methods, can introduce noise into the objective function and its derivatives
\cite[pp.~44,~110]{betts2010pmo}.
Nevertheless, estimates of the integration error are useful to evaluate the need
for mesh refinement or alternative integration schemes.

Let us denote by $\im$ one step of the integration method,
\begin{equation}
  \label{eq:integ_method}
  \xp_{k+1} =
  \im(\xp_k,\tau_k,\tau_{k+1}, u, \theta,f),
\end{equation}
i.e., the method applied to the solution of the differential equation defined
by $f$, with the input function $u$, starting with the value $\xp_k$ at the
time $\tau_k$ and returning the value $\xp_{k+1}$ corresponding to $\tau_{k+1}$.
With Euler's method, for example, we would have
\begin{equation}
  \im(\xp_k,\tau_k,\tau_{k+1}, u, \theta,f) = 
  \xp_k + (\tau_{k+1} - \tau_k) f\big(\xp_k,u(\tau_k), \theta\big).
\end{equation}
The output-error method is then
\begin{subequations}
  \label{eq:oem_ivp}
  \begin{alignat}{4}
    \label{eq:oem_ivp_obj}
    \maximize_{\theta\in\reals^\npar} & \quad&&
    \sum_{\substack{{k=1}\\ \tau_i = \tm_k}}^N \elle(z_k - \yp_i, \theta)\\
    \label{eq:oem_ivp_cinit}
    \subjectto &&&
    \xp_1 = x_0(\theta) \\
    \label{eq:oem_ivp_cxdot}
    &&&
    \xp_{k+1} =
    \im(\xp_k,\tau_k,\tau_{k+1}, u, \theta,f), 
    &\qquad&  k=1,\dotsc,m-1\\
    \label{eq:oem_ivp_cy}
    &&&\yp_k = g\big(\xp_k, u(\tau_k), \theta\big),
    &&  k=1,\dotsc,m.
  \end{alignat}
\end{subequations}%
\noeqref{eq:oem_ivp_cxdot,eq:oem_ivp_cy, 
         eq:oem_ivp_cinit,eq:oem_ivp_obj}%
Eqs.~\eqref{eq:oem_ivp_cinit}--\eqref{eq:oem_ivp_cy} are explicit
and can be calculated one after the other, iteratively.

When the $f$, $g$ and $\elle$ functions are continuous and differentiable
the problem is amenable to solution with Newton's method.
The gradients can be obtained by finite differences 
\cite[Sec.~4.8]{jategaonkar2015fvs}, automatic differentiation or symbolic
differentiation of the model functions.
The resulting nonlinear programming problem can then be readily solved with a
wide range of available software packages like the open-source IPOPT from
the COIN-OR initiative \cite{wachter2006iip}.
If the Gaussian distribution is assumed for the measurement error in the
form of \eqref{eq:pe_gaussian}, \eqref{eq:ml_gauss_known_cov} or 
\eqref{eq:ml_gauss_known_cov_diag}, 
nonlinear least-squares optimization methods such as the Gauss--Newton
or Levenberg--Marquardt algorithms can be used to obtain the solution
\cite[Sec.~4.6 and~4.13]{jategaonkar2015fvs}.
The problem is dense but relatively small-sized, usually with up to hundreds of
decision variables which are the elements of the unknown parameter vector
$\theta$.

However, it should be noted that the numerical solution of the differential
equation might diverge or deviate far from the measurements for parameter
values far from the optimum.
This can, in turn, lead to divergence of the optimization or convergence to
suboptimal local maxima.
Furthermore, in unstable, marginally stable, poorly damped or systems with
slow dynamics the errors in the integration method can accumulate or be
amplified by the system dynamics, degrading the estimates.

\subsection{Multiple shooting implementation}
\label{sec:multiple_shooting}
In the multiple shooting implementation [\citenum{bock1981nti}; 
\citenum{bock1983rap}; \citenum{jategaonkar2015fvs}, Sec.~9.12] of the 
output-error method, a numerical integration scheme is used as in the single
shooting approach, but divided into multiple segments with independent initial
states which are appended to the decision variables of the optimization.
Equality constraints are then included to enforce continuity of the state
solution across segments.

By the choice of a grid
\begin{align}
  \label{eq:multiple_shooting_grid}
  0 = t^{(1)} < t^{(2)} < \cdots < t^{(\ns + 1)} = T.
\end{align}
the experiment interval is divided into $\ns$ segments, each of which is
further subdivided by a finer mesh of $m_k$ points for numerical integration,
\begin{equation}
  \label{eq:multiple_shooting_mesh}
  t^{(k)} = 
  \tau^{(k)}_1 < \tau^{(k)}_2 < \cdots < \tau^{(k)}_{m_k} 
  = t^{(k + 1)},
  \qquad  k = 1,\dotsc,\ns.
\end{equation}
The vector  $\xi$ of decision variables of the optimization is then 
augmented with the state at the beginning of each segment after the first,
\begin{align}
  \label{eq:multiple_shooting_decision}
  \xi =
  \begin{bmatrix}
    \theta &
    \xp^{(2)}_1 &
    \xp^{(3)}_1 &
    \cdots &
    \xp^{(\ns)}_1
  \end{bmatrix}
\end{align}
and the state equations \eqref{eq:state_dynamics} are integrated using the
chosen numerical scheme along the mesh of each segment, independently.

Using the same notation for the integration scheme as in 
\eqref{eq:integ_method}, the multiple shooting approach for the
output-error method is
\begin{subequations}
  \label{eq:oem_ms}
  \begin{alignat}{5}
    \label{eq:oem_ms_obj}
    \maximize_{\quad\theta,\xp^{(2)}_0, \dotsc, \xp^{(\ns)}_0} 
    & \quad&&
    \sum_{\substack{{k=1}\\ \tau_i^{(j)} = \tm_k}}^N 
    \elle\big(z_k - \yp_i^{(j)}, \theta\big)\\
    \label{eq:oem_ms_cinit}
    \subjectto\quad &&&
    \xp_1^{(1)} = x_0(\theta) \\
    \label{eq:oem_ms_cxdot}
    &&&
    \xp_{k+1}^{(i)} =
    \im(\xp_k^{(i)},\tau_k^{(i)},\tau_{k+1}^{(i)}, u, \theta,f), 
    &\qquad&  k=1,\dotsc,m_i-1;
    &\qquad&  i=1,\dotsc,\ns\\
    \label{eq:oem_ms_cy}
    &&&\yp_k^{(i)} = g\big(\xp_k^{(i)}, u(\tau_k^{(i)}), \theta\big),
    &&  k=1,\dotsc,m_i;
    &&  i=1,\dotsc,\ns\\
    \label{eq:oem_ms_cont}
    &&& \xp_{m_k}^{(i)} - \xp_1^{(i+1)} = 0
    && i=1,\dotsc,\ns-1.
  \end{alignat}
\end{subequations}%
\noeqref{eq:oem_ms_cxdot,eq:oem_ms_cy,
         eq:oem_ms_cinit,eq:oem_ms_obj,eq:oem_ms_cont}%
Eqs.~\eqref{eq:oem_ms_cinit}--\eqref{eq:oem_ms_cy} are explicit
and can be calculated one after the other, iteratively.
In addition, the iterations of each segment $i$ are independent, permitting
a considerable speedup when using modern computer hardware with parallel
capabilities.
Finally, we note that \eqref{eq:oem_ms_cont} are the continuity constraints,
requiring that the final point of each segment coincide with the initial
point of the next.
They are, generally, nonlinear functions of the decision variables even for
linear dynamical systems.
The optimization problem resulting from \eqref{eq:oem_ms} is larger, due to the
inclusion of the additional variables and constraints, but it is sparse.

This method is far more robust against divergence of the numerical solution
to the differential equation than single shooting.
As each segment is kept short, even when for inadequate parameter values, 
the solution does not deviate much from the first point of the segment.
Furthermore, from the very inverse nature of the inverse problem, more 
information on the states is usually available than on the parameters,
allowing for better starting values for the optimization.

The continuity constraints are enforced by the optimization solver with
a small tolerance which can offset the integration errors accumulated
on unstable, marginally stable or poorly damped systems.
This allows the method to be used without modifications on this
important class of systems, yielding good results.
Furthermore, the expansion of the search domain gives more degrees of freedom
to the optimization solver, which can now explore search directions which
violate the continuity constraints.
This endows the method with better numerical and convergence properties,
allowing it to overcome local minima which would otherwise capture the
single shooting implementation.

\subsection{Collocation implementation}
The collocation approach for formulating the output-error method consists of
adding all (or most) state points used by the
integration scheme as decision variables of the optimization and encoding
the integration method as equality constraints.
Instead of iterating the integration scheme, the integration error of the
candidate solution is evaluated and passed on to the optimization routine,
which drives it to zero at the same time it works to maximize the objective.
Because of this, implicit integration schemes are a natural choice for
collocation methods.

The method is very general with many variations, so we will exemplify it here 
with a family of Lobatto methods which includes the trapezoidal method.
A very thorough reference on collocation methods for parameter estimation, with
implementation details and examples is \citet{betts2010pmo}.
Additional information can also be found in the works of \citet{betts2003lsp}
and \citet{williams2005ope}.

Like in the multiple shooting, the experiment interval is divided into a 
grid of $\ns$ segments \eqref{eq:multiple_shooting_grid}, each of which is
further subdivided into a finer mesh \eqref{eq:multiple_shooting_mesh} of $m_k$
points,
including the endpoints.
The vector  $\xi$ of decision variables of the optimization problem is then 
augmented with the state at all mesh points,
\begin{align}
  \label{eq:collocation_decision}
  \xi =
  \begin{bmatrix}
    \theta &
    \xp^{(1)}_1 &
    \cdots &
    \xp^{(1)}_{m_1} &
    \xp^{(2)}_2 &
    \cdots &
    \xp^{(2)}_{m_2} &
    \xp^{(3)}_2 &
    \cdots &
    \xp^{(3)}_{m_3} &
    \cdots &
    \xp^{(\ns)}_{m_\ns}
  \end{bmatrix}.
\end{align}
Notice that the end of each segment coincides with the start of the next,
but only one decision variable is needed to represent it.
To simplify notation in what follows, we will use both $\xp_{m_i}^{(i)}$ and
$\xp_{1}^{(i+1)}$ to refer to these variables.

To derive the method, we first note that both sides of the differential equation
\eqref{eq:state_dynamics} can be integrated, leading to the integral form of the
system equations:
\begin{equation}
  \label{eq:state_dynamics_int_form}
  x(t_b) - x(t_a) = \int_{t_a}^{t_b} f\big(x(\tau),u(\tau),\theta\big)\,\dd\tau 
  \qquad\forall t_a,t_b\in[0,T].
\end{equation}
This integral relation holds for any two time values in the experiment interval.
A way of approximating this relation is to build a polynomial approximation
$\hat f(t)$ of the function $f$ at the mesh points and enforce
\eqref{eq:state_dynamics_int_form} to hold for $\hat f$ and any pair 
$t_a$, $t_b$ in the mesh.

Let $\lambda_k^{(i)}(t)$ denote the $k$-th Lagrange basis polynomial for the 
mesh of the $i$-th segment:
\begin{equation}
  \label{eq:lagrange_basis}
  \lambda_k^{(i)}(t) = \prod_{\substack{1 \leq j \leq m_i \\ j\neq k}}
  \frac{t-\tau_j^{(i)}}{\tau_k^{(i)} - \tau_j^{(i)}}.
\end{equation}
This basis is defined such that
\begin{align}
  \label{eq:lagrange_basis_prop}
  \lambda_k^{(i)}(\tau_k^{(i)}) & =1, &
  \lambda_k^{(i)}(\tau_j^{(i)}) & = 0, \quad \forall j\neq k,
\end{align}
simplifying the construction of a polynomial interpolant of $f$ at the grid
points:
\begin{align}
  \label{eq:fhat_interp}
  \hat f^{(i)}(t) = \sum_{k=1}^{m_i} \lambda_k^{(i)}(t) 
  f\big(\xp_k^{(i)}, u(\tau_k^{(i)}),\theta\big).
\end{align}
If we apply the integral form of the system equation 
\eqref{eq:state_dynamics_int_form}
to $\hat f$, we obtain the collocation constraints for this family of 
Lobatto methods:
\begin{align}
  \label{eq:colloc_constraint}
  \xp_{k+1}^{(i)} - \xp_k^{(i)} = 
  \int_{\tau_k^{(i)}}^{\tau_{k+1}^{(i)}}\hat f^{(i)}(\tau)\,\dd\tau =
  \sum_{j=1}^{m_i} f\big(\xp_j^{(i)}, u(\tau_j^{(i)}),\theta\big)
  \int_{\tau_k^{(i)}}^{\tau_{k+1}^{(i)}} \lambda_j^{(i)}(\tau) \,\dd\tau.
\end{align}
We note that the last integral on the right-hand side of 
\eqref{eq:colloc_constraint}
depends only on the location of the mesh points and can be calculated beforehand
\begin{align}
  \label{eq:coll_integral_defn}
  C_{j,k}^{(i)} := 
  \int_{\tau_k^{(i)}}^{\tau_{k+1}^{(i)}} \lambda_j^{(i)}(\tau) \,\dd\tau.
\end{align}
Furthermore, as the integral of a polynomial it has a simple analytic solution.
The mesh points must be carefully chosen to avoid Runge's phenomenon and 
guarantee a low bound on the solution error.
A good general choice are the Legendre--Gauss--Lobatto nodes of the segment
\citep[Sec.~3]{williams2005ope}.

The collocation-based output-error method then amounts to the following 
optimization problem
\begin{subequations}
  \label{eq:oem_coll}
  \begin{alignat}{5}
    \label{eq:oem_coll_obj}
    \maximize_{
      \theta,\xp^{(1)}_1,\dotsc,\xp^{(1)}_{m_1},
      \xp^{(2)}_2,\dotsc,\xp^{(\ns)}_{m_\ns}
    } 
    & \quad&&
    \sum_{\substack{{k=1}\\ \tau_i^{(j)} = \tm_k}}^N 
    \elle\big(z_k - \yp_i^{(j)}, \theta\big)\\
    \label{eq:oem_coll_cinit}
    \subjectto\qquad\; &&&
    0 = \xp_1^{(1)} - x_0(\theta) \\
    \label{eq:oem_coll_cxdot}
    &&&
    0=\xp_{k+1}^{(i)} - \xp_k^{(i)} - 
    \sum_{j=1}^{m_i} f\big(\xp_j^{(i)},u(\tau_j^{(i)}),\theta\big)C^{(i)}_{j,k},
    &\quad&  k=1,\dotsc,m_i-1;
    &\quad&  i=1,\dotsc,\ns\\
    \label{eq:oem_coll_cy}
    &&&\yp_k^{(i)} = g\big(\xp_k^{(i)}, u(\tau_k^{(i)}), \theta\big),
    &&  k=1,\dotsc,m_i;
    &&  i=1,\dotsc,\ns.
  \end{alignat}
\end{subequations}%
\noeqref{eq:oem_coll_cxdot,eq:oem_coll_cy,
         eq:oem_coll_cinit,eq:oem_coll_obj}%
Of the three approaches presented in this section, collocation leads to the 
largest optimization problem, but also the most sparse.
Both the objective function and the constraints are very simple and can be 
parallelly evaluated.
Each segment's mesh and length is usually kept small, such that there is little 
variation of the state and inputs within it, which makes the linearization
of the constraints by sequential quadratic programming solvers reasonable
approximations.
When most states are measured, good initial values of the states can be used 
for the optimization, improving convergence even starting with inadequate
parameter values.

The same considerations made on the end of Sec.~\ref{sec:multiple_shooting}
about the multiple shooting approach also apply here, even to a higher degree.
Divergence of the numerical integration does not make sense unless the whole
vector of decision variables diverges.
The error tolerance on the constraint violation prevents integration error
from accumulating even in a single integration step.
The segments are even smaller and the collocation constraint is closer to
a linear function, as the function $f$ is not applied recursively into
itself.
This makes the problem more amenable to solution with Newton's method, which
employs quadratic approximations for the objective and linear approximations
for the constraints.

\section{Examples}
\label{sec:examples}
We illustrate the collocation-based output-error method with three experiments:
a linear model parameter estimation of the short period dynamics from a
simulated unstable aircraft;
a nonlinear longitudinal model estimation of a HFB-320 Hansa Jet from 
experimental data collected by the DLR;
and a high order (24 states) black-box linear model estimation from an
experimental ground vibration test of a unmanned flexible flying wing aircraft.
The examples were chosen to demonstrate the robustness of the collocation 
approach to poor start parameters; its applicability, without modifications,
to parameter estimation in unstable systems; and its use in a challenging 
large-scale problem, with several states and many samples, typical of 
flexible vehicle parameter estimation.
The source code of the analyses performed for this article were released as
open-source software\footnote{%
  Available for download at \url{https://github.com/dimasad/aviation-2019-code}
}.

The estimation routines were implemented in the open-source CEA Control and
Estimation Library (CEACoEst)%
\footnote{%
  Available for download at \url{https://github.com/cea-ufmg/ceacoest}
}, under development by
the author.
It is written in the Python programming language and uses the open-source
large-scale interior point nonlinear optimization solver
IPOPT \cite{wachter2006iip}, part of the COIN-OR initiative.
The solver, in turn, uses the efficient sparse linear system solvers of the 
HSL Mathematical Software Library\footnote{%
  HSL. A collection of Fortran codes for large scale scientific computation. 
  \url{http://www.hsl.rl.ac.uk/}
} to obtain the search direction of Newton's method.
The HSL and OpenBLAS \cite{wang2013augem} libraries in the implementation
used were compiled with the OpenMP parallel computing framework 
\cite{openmp2013omp} enabled, exploiting the parallelism inherent in the
collocation formulation.

\subsection{HFB-320 Hansa Jet longitudinal motion}
Experimental data collected from a HFB-320 Hansa Jet aircraft by the DLR
are availabe with the supplemental material\footnote{%
  Freely available for download at 
  \url{https://arc.aiaa.org/doi/suppl/10.2514/4.102790}
} of the book by \citet{jategaonkar2015fvs}.
It consists of an elevator 3-2-1-1 input followed by a pulse.

A nonlinear model was estimated with 4 states, the airspeed $V$, 
angle of attack $\alpha$, pitch angle $\vartheta$, pitch rate $q$; 2~inputs,
the elevator deflection $\delta_e$ and the thrust force $T$; and 
7 measurements, $\tilde V$, $\tilde\alpha$, $\tilde\vartheta$, $\tilde q$ 
and the pitch acceleration $\tilde a_q$, longitidinal acceleration $\tilde a_x$,
and vertical acceleration $\tilde a_z$.
A total of 28 unknown parameters $\theta$ were estimated: the dimensionless
stability and control derivatives 
$C_{D_0}$, $C_{D_V}$, $C_{D_\alpha}$, $C_{L_0}$, $C_{L_V}$, $C_{L_\alpha}$, 
$C_{m_0}$, $C_{m_V}$, $C_{m_\alpha}$, $C_{m_q}$, and $C_{m_{\delta_e}}$;
the measurement biasses $b_{q}$, $b_{a_q}$, $b_{a_x}$, and $b_{a_z}$; 
the measurement noise standard deviations $\sigma_{V}$, $\sigma_{\alpha}$,
$\sigma_{\vartheta}$, $\sigma_{q}$, $\sigma_{a_q}$, $\sigma_{a_x}$, and 
$\sigma_{a_z}$; and the initial states $V_0$, $\alpha_0$, $\theta_0$, and
$q_0$.

The function $f$ of the dynamics is defined by the following relations:
\begin{align}
  \dot V &= -\frac{\rho S}{2 m} V^2
  \big(C_{D_0} + C_{D_V}(\tfrac{V}{\Vref} - 1) + C_{D_\alpha}\alpha\big) 
  +\frac{T}{m} \cos(\alpha+\epsilon_T) - g_0\sin(\theta - \alpha) \\
  \dot \alpha &= -\frac{\rho S}{2 m} V
  \big(C_{L_0} + C_{L_V}(\tfrac{V}{\Vref} - 1) + C_{L_\alpha}\alpha\big) 
  -\frac{T}{mV} \sin(\alpha + \epsilon_T) + \frac{g_0}{V}\cos(\theta - \alpha)
  + q
  \\
  \dot \vartheta &= q \\
  \dot q &= 
  \frac{\rho S\bar c}{2 I_y} V^2
  \left(
  C_{m_0} + C_{m_V}(\tfrac{V}{\Vref} - 1) + C_{m_\alpha}\alpha
  + C_{m_q} \frac{\bar c q}{2 V} + C_{m_{\delta_e}}\delta_e
  \right) 
  +\frac{\ell_T}{I_y}T.
\end{align}
Similarly, the function $g$ of the outputs is defined by
\begin{align}
  \tilde V &= V \\
  \tilde \alpha &= \alpha \\
  \tilde\vartheta &= \vartheta\\
  \tilde q &= q + b_q\\
  \tilde a_q &= 
  b_{a_q} + 
  \frac{\rho S\bar c}{2 I_y} V^2
  \left(
  C_{m_0} + C_{m_V}(\tfrac{V}{\Vref} - 1) + C_{m_\alpha}\alpha
  + C_{m_q} \frac{\bar c q}{2 V} + C_{m_{\delta_e}}\delta_e
  \right) 
  +\frac{\ell_T}{I_y}T \\
  \tilde a_x &= 
  b_{a_x} + 
  \frac{\rho S}{2 m} V^2
  \left[
    \sin\alpha
    \big(C_{L_0} + C_{L_V}(\tfrac{V}{\Vref} - 1) + C_{L_\alpha}\alpha\big)
    -\cos\alpha
    \big(C_{D_0} + C_{D_V}(\tfrac{V}{\Vref} - 1) + C_{D_\alpha}\alpha\big)
  \right]  
  +\frac{T}{m} \cos(\epsilon_T)
  \\
  \tilde a_z &= 
  b_{a_z} + 
  \frac{\rho S}{2 m} V^2
  \left[
    -\cos\alpha
    \big(C_{L_0} + C_{L_V}(\tfrac{V}{\Vref} - 1) + C_{L_\alpha}\alpha\big)
    -\sin\alpha
    \big(C_{D_0} + C_{D_V}(\tfrac{V}{\Vref} - 1) + C_{D_\alpha}\alpha\big)
  \right]  
  -\frac{T}{m} \sin(\epsilon_T).
\end{align}
Finally, the $\elle$ log-density associated with Gaussian noise 
\eqref{eq:pe_gaussian} was used, with a diagonal $R(\theta)$ consisting of the 
measurement variances $\sigma_V^2$ to $\sigma_{a_z}^2$.

\begin{figure}
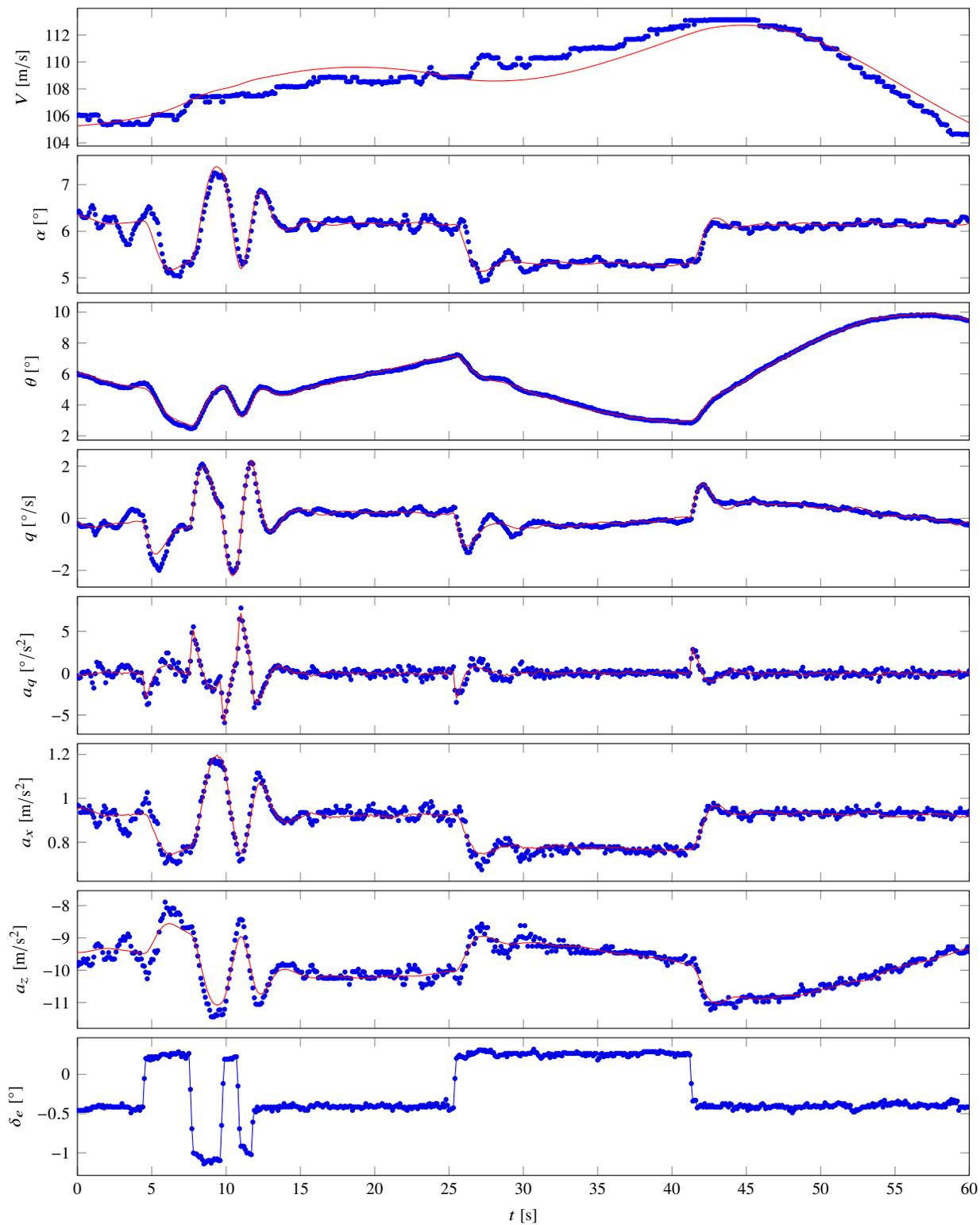

  \centering
  \inputtikzpicture{hfb320}
  \caption{%
    Estimation results for the HFB320 data.
    The marks are the measured values and the solid lines are the estimated
    model output.
    The elevator input for the manuever is the bottommost plot.
  }
  \label{fig:hfb320}
\end{figure}

\begin{table}
  \centering
  \pgfplotstabletypeset[
    col sep=&,row sep=\\,
    columns/parameter/.style={string type},
    columns/optimal/.style={fixed, fixed zerofill, precision=4, dec sep align},
  ]{
    parameter & lower starting value & upper starting value & optimal \\
    {$C_{D_0}$} & 0 & 0.5& 5.802198227121584118e-02 \\
    {$C_{D_V}$} & -0.5 & 0.5& -3.160934509655806413e-02 \\
    {$C_{D_\alpha}$} & 0 & 1& 2.453480135927716133e-01 \\
    {$C_{L_0}$} & 0 & 2& 1.808149665501641579e-01 \\
    {$C_{L_V}$} & -2 & 2& 2.011780626599658228e-01 \\
    {$C_{L_\alpha}$} & 0 & 10& 3.090429255239008466e+00 \\
    {$C_{m_0}$} & 0 & 0.5& 1.183601781319232743e-01 \\
    {$C_{m_V}$} & 0 & 0.5& 1.366124982685849344e-02\\
    {$C_{m_\alpha}$} & -5 & 1& -9.940542701262951031e-01\\
    {$C_{m_q}$} & -50 & 0& -2.865165518164920755e+01 \\
    {$C_{m_{\delta_e}}$} & -10 & 0& -1.471386926974762144e+00\\
  }
  \caption{Limits of the uniform distribution from which the starting
    parameters of the HFB-320 example were drawn, together with their 
    optimal values.
  }
  \label{tab:hfb320}
\end{table}

The estimation was performed a total of \num{1000} times with random starting
values for the stability and control derivatives, drawn from uniform
distributions whose intervals are shown in Tab.~\ref{tab:hfb320}.
The starting value for the bias parameters were zero and for the noise standard
deviations was one.
The mesh states were initialized with the measured values.
Of all starting points of the optimization, \SI{98.6}{\%} converged to 
the global optimum whose parameter values are listed in the last column of 
Tab.~\ref{tab:hfb320} and model outputs are shown in Fig.~\ref{fig:hfb320}.

We note that the optimization converges even for very bad values of $\theta$
which pose problems for the single shooting implementation.
Particularly, if we set all aerodynamic derivatives to zero, there is no
lift, drag or aerodynamic pitching moment and the model plummets down, its
initial-value problem simulation diverging.
However, the collocation-based estimation manages to reach the optimal solution.

\subsection{Simulated unstable aircraft short period}

\begin{figure}[t]
  \centering
  \inputtikzpicture{uac}
  \caption{%
    Estimation results for the simulted unstable aircraft data.
    The marks are the measured values and the solid lines are the estimated
    model output.
    The elevator input for the manuever is the bottommost plot.
  }
  \label{fig:uac}
\end{figure}

To demonstrate that the collocation-based output-error method can be used
without modifications to estimate unstable dynamics, we apply it on data
from an unstable simulated aircraft.
The data was used by
\citet[Sec.~9.16.1]{jategaonkar2015fvs} to compare various methods for unstable
aircraft system identification.
It should be noted that, for this system, the initial-value problem approach
for the output-error method obtains very poor estimates.

A linear model of the short period dynamics was used with 2 states,
the vertical velocity $w$ and pitch rate $q$; only 1 input, the elevator 
deflection $\delta_e$; and 3 outputs, $\tilde w$, $\tilde q$, and the vertical
acceleration $\tilde a_z$.
A total of 11 parameters were estimated, the dimensional stability and 
control derivatives $Z_w$, $Z_q$, $Z_{\delta_e}$, $M_w$, $M_q$, and 
$M_{\delta_e}$; the measurement noise standard deviations $\sigma_{w}$,
$\sigma_{q}$, and $\sigma_{a_z}$; and the initial states $w_0$ and $q_0$.

The system dynamics is defined by
\begin{align}
  \dot w &= Z_w w + (U_0 + Z_q) q + Z_{\delta_e}\delta_e, &
  \dot q &= M_w w + M_q q + M_{\delta_e}\delta_e,
\end{align}
and its outputs by
\begin{align}
  \tilde w &= w, &
  \tilde q &= q, &
  \tilde a_z&= Z_w w + Z_q q + Z_{\delta_e}\delta_e.
\end{align}
The $\elle$ log-density associated with Gaussian noise 
\eqref{eq:pe_gaussian} was used, with a diagonal $R(\theta)$ consisting of the 
measurement variances $\sigma_w^2$ to $\sigma_{a_z}^2$.
The starting mesh states were the measured values, the stability and
control derivatives were zero, and the measurement noise standard deviations
were one.
The estimation model outputs for the collocation-based output-error method
are shown in Fig.~\ref{fig:uac}.

\subsection{Black-box high-order ground vibration test}
For the estimation of models of the rigid-body flight dynamics of aircraft,
most states are measured directly and the model structure is well-known.
The estimation of flexible aircraft dynamics poses additional challenges as the
structural states are not usually measured and the model order is much larger.
Many of these issues also arise in the system identification of structural 
models from ground vibration tests.
To illustrate the application of the collocation-based output-error method to
this class of problems, it was applied to accelerometer data from a
ground vibration test reported by  \citet{gupta2016gvt}, part of the 
Performance Adaptive Aeroelastic Wing program\footnote{%
  \url{http://www.paaw.net/}
}.

A single-input single-output model with 24 states was estimated for the 
force input $u$ and acceleration output $y$.
The dynamics were represented in the state-space modal form:
\begin{align}
  \dot x_{1a} &= \sigma_1 x_{1a} + \omega_1 x_{1b} + b_1 u \\
  \dot x_{1b} &= -\omega_1 x_{1a} + \sigma_1 x_{1b}\\
  &\;\;\vdots \\
  \dot x_{12a} &= \sigma_{12} x_{12a} + \omega_{12} x_{12b} + b_{12} u \\
  \dot x_{12b} &= -\omega_{12} x_{12a} + \sigma_{12} x_{12b},
\end{align}
such that the system poles are located at $\sigma_i \pm j\omega_i$.
The system output equation is
\begin{equation}
  y = c_{1a} x_{1a} + c_{1b} x_{1b} + \dots + c_{12a} x_{12a} + c_{12b} x_{12b}.
\end{equation}
The $\elle$ log-density associated with Gaussian noise 
\eqref{eq:pe_gaussian} was used, with $R(\theta) = \sigma^2_y$.
The unknown parameter vector consists of the pole locations $\sigma_i$ and
$\omega_i$, the measurement standard deviation $\sigma_y$, and the 
coefficients $b_i$, $c_{ia}$, and $c_{ib}$.

\begin{table}
  \centering
  \pgfplotstabletypeset[
    col sep=&,row sep=\\,
    columns/a/.style={column name=$\omega_1$},
    columns/b/.style={column name=$\omega_2$},
    columns/c/.style={column name=$\omega_3$},
    columns/d/.style={column name=$\omega_4$},
    columns/e/.style={column name=$\omega_5$},
    columns/f/.style={column name=$\omega_6$},
  ]{
    a & b & c & d & e & f \\
    199.06 & 186.17 & 178.44 & 172.64 & 144.37 & 121.47\\
  }
  
  \pgfplotstabletypeset[
    col sep=&,row sep=\\, fixed zerofill, precision=2,
    columns/a/.style={column name=$\omega_7$},
    columns/b/.style={column name=$\omega_8$},
    columns/c/.style={column name=$\omega_9$},
    columns/d/.style={column name=$\omega_{10}$},
    columns/e/.style={column name=$\omega_{11}$},
    columns/f/.style={column name=$\omega_{12}$},
  ]{
    a & b & c & d & e & f \\
    96.515 & 59.572 & 58.567 & 59.029 & 48.989 & 48.901 \\
  }
  \caption{%
    Starting values for the pole frequencies in the ground vibration 
    test example.
  }
  \label{tab:gvt}
\end{table}

The data collected contains high-frequency components which we do not wish
to consider, so both the input and output were filtered with band-pass filters
with a passband of \SIrange{6}{32}{Hz}, composed of cascaded 5th-order
Butterworth filters.
The data was then decimated by 20 for the estimation, dropping the sampling
rate from \SI{2}{kHz} to \SI{100}{Hz}.
The starting value for the parameters was zero for the $\sigma_i$ and mesh 
states, one for the $c_{ia}$, $c_{ib}$ and $b_i$.
The starting values of the $\omega_i$ were obtained from the poles of a model
identified with the N4SID algorithm which are shown in Tab.~\ref{tab:gvt}.

\begin{figure}
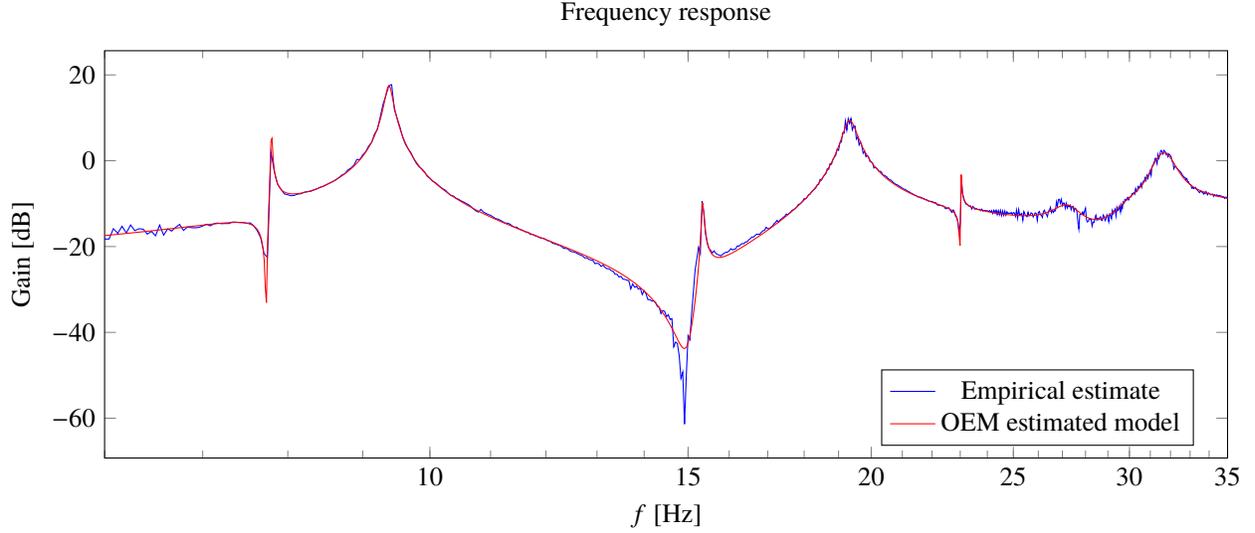

  \centering
  \inputtikzpicture{gvt_freq}
  \caption{%
    Frequency response magnitude of the empirical transfer function estimate
    and of the OEM-estimated model for the ground vibration test.
  }
  \label{fig:gvt_freq}
\end{figure}

\begin{figure}
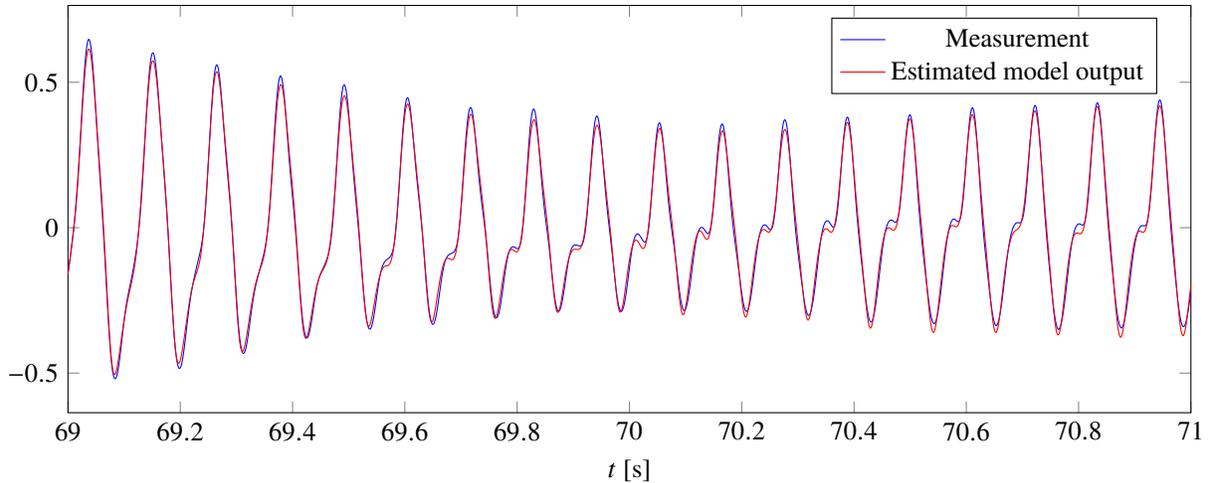

  \centering
  \inputtikzpicture{gvt_ysim}
  \caption{%
    Portion of the time-response of the identified model in the ground
    vibration test example.
  }
  \label{fig:gvt_ysim}
\end{figure}

The frequency response of the identified model is shown in 
Fig.~\ref{fig:gvt_freq}, together with that of the empirical transfer function
estimate from the test data.
A portion of the estimated model outputs shown against the data can also
be seen in Fig.~\ref{fig:gvt_ysim}.

\section{Conclusions and future work}
\label{sec:conclusion}
The collocation-based output-error method for parameter estimation, although
seldomly used in aircraft system identification, better exploits the nature of
the inverse problem and has advantages over the ubiquitous single shooting 
implementations:
applicability to a wider class of systems, such as unstable, marginally
stable or poorly damped systems;
better numerical and convergence properties such as robustness against poor
parameter values;
parallelizable formulation of the objective function and constraints which
can benefit from modern computer hardware and capabilities such
as single instruction, multiple data (SIMD), multicore central processing
units (CPUs), graphics processing units (GPUs) and computer clusters.
These properties can help overcome the challenges in demanding applications
like the identification of flexible aircraft dynamics which pose
difficulties to the single shooting method.

We note, however, that the main principle behind the collocation method is
fairly general: augmenting the optimization decision variables with the states
and encoding their dynamics as equality constraints.
This overall technique can also be applied to other more general parameter
estimation methods as well, notably the filter-error method, lending it
similar improvements as for the output-error method.

\bibliography{bib}

\end{document}